\documentclass[prl,twocolumn,showpacs]{revtex4}

\usepackage{graphicx}
\usepackage{dcolumn}
\usepackage{bm}
\usepackage{amsmath}
  
\begin{document}

\title{A More Accurate Generalized Gradient Approximation for Solids} 

\author{Zhigang Wu}
\email{z.wu@gl.ciw.edu}
\affiliation{Geophysical Laboratory, Carnegie Institution of Washington, Washington, DC 20015, USA}
\author{Ronald E. Cohen}
\affiliation{Geophysical Laboratory, Carnegie Institution of Washington, Washington, DC 20015, USA}

\date{\today}

\begin{abstract}
We present a new nonempirical density functional generalized gradient 
approximation (GGA) that gives significant improvements for lattice 
constants, crystal structures, and metal surface energies over the most 
popular Perdew-Burke-Ernzerhof (PBE) GGA. The new functional is based on 
a diffuse radial cutoff for the exchange-hole in real space, and the 
analytic gradient expansion of the exchange energy for small gradients. 
There are no adjustable parameters, the constraining conditions of PBE are 
maintained, and the functional is easily implemented in existing codes.

\end{abstract}


\maketitle

Kohn-Sham density functional theory (DFT) \cite{dft1, dft2} makes it possible to solve many-electron 
ground-state problems efficiently and accurately. The DFT is exact if the exchange-correlation 
(XC) energy $E_{\rm XC}$ were known exactly, but there is no tractable exact expressions of 
$E_{\rm XC}$ in terms of electron density. Numerous attempts have been made to approximate
$E_{\rm XC}$, starting with the local (spin) density (LSD) approximation (LDA), which is still 
widely used. The generalized gradient approximations (GGAs) \cite{pw86,pw91,pbe} are semilocal, 
seeking to improve upon LSD. Other more complicated approximations are often orbital-dependent 
or/and nonlocal. They suffer from computational inefficiency; it is much harder to treat them 
self-consistently and to calculate energy derivative quantities.

The XC energy of LSD and GGAs are
\begin{equation}
E^{\rm LSD}_{\rm XC}[n_{\uparrow}, n_{\downarrow}] =  
\int n \epsilon_{\rm XC}^{\rm unif}(n_{\uparrow}, n_{\downarrow}) d^{3}r, 
\label{eq1}
\end{equation}
and
\begin{equation}
E^{\rm GGA}_{\rm XC}[n_{\uparrow}, n_{\downarrow}] =  
\int f(n_{\uparrow}, n_{\downarrow}, \nabla n_{\uparrow}, \nabla n_{\downarrow}) d^{3}r, 
\label{eq2}
\end{equation}
respectively. Here the electron density $n = n_{\uparrow} + n_{\downarrow}$, and 
$\epsilon_{\rm XC}^{\rm unif}$ is the XC energy density for the uniform electron gas.
LSD is the simplest approximation, constructed from uniform electron gas, and very
successful for solids, where the valence electron densities vary relatively more slowly
than in molecules and atoms, for which GGAs \cite{pbe,perdew92} achieved a great improvement
over LSD. It is well known that LSD underestimates the equilibrium 
lattice constant $a_{0}$ by 1-3\%, and some properties such as ferroelectricity are 
extremely sensitive to volume. When calculated at the LSD volume, the ferroelectric instability 
is severely underestimated \cite{cohen90,cohen92,singh92}. 
On the other hand, GGAs tend to expand lattice constants. They well predict 
correct $a_{0}$ for simple metals, such as Na and K \cite{perdew92}, however for other 
materials they often overcorrect LSD by predicting $a_{0}$ 1-2\% bigger \cite{star04}
than experiment. Predicting lattice constants more accurately than LSD remains a
tough issue, even for state-of-the-art meta-GGAs; nonempirical TPSS \cite{tpss} only 
achieves moderate improvement over PBE, while empirical PKZB \cite {pkzb} is worse than PBE. 
GGAs are especially poor for ferroelectrics, e.g., PBE \cite{pbe} predicts the volume and 
strain of relaxed tetragonal PbTiO$_{3}$ more than 10\% and 200\% too large, 
respectively \cite{wu04}, and other GGAs \cite{pw91,revpbe,rpbe} are even worse, 
as seen in Table \ref{tab1}. Another more complicated functional, the nonlocal weighted 
density approximation (WDA) is also unsatisfactory for this case \cite{wu04}. 
To compute these properties correctly, people often constrain volumes at their experimental 
values $V_{\rm expt}$. However, $V_{\rm expt}$ is not available for predicting new 
materials, certain properties are still wrong even at $V_{\rm expt}$; theoretically it 
is more satisfactory to do calculations without any experimental data adjustments.


\begin{table}[tbp]
\caption{\label{tab1} 
Calculated equilibrium volume $V_{0}$ (\AA$^{3}$) and strain (\%) of tetragonal 
PbTiO$_{3}$ for various GGAs comparing with experimental data at low
temperature \cite{mabud}.}

\begin{ruledtabular}
\begin{tabular}{lccccc}
         & PW91   & PBE    & revPBE & RPBE  & Expt.         \\
\hline
 $V_{0}$ & 70.78  & 70.54  & 74.01  & 75.47 & 63.09         \\
 strain  & 24.2   & 23.9   & 28.6   & 30.1  & 7.1           \\

\end{tabular}
\end{ruledtabular}

\end{table}


In order to study finite temperature properties, e.g., ferroelectric phase transitions, 
effective Hamiltonian and potential models, which are used in molecular dynamics (MD) or Monte 
Carlo (MC) simulations, have been developed with parameters fitted
to first-principles results. When the model parameters are fitted to 
LSD data, these simulations greatly underestimate the phase transition temperatures $T_{\rm c}$ 
at ambient pressure \cite{zhong94,marcelo04,krakauer99}, 
and overestimate if fitted to the GGA results \cite{marcelo04}.  
A simple but more accurate approximation for XC energy is necessary.


\begin{figure}[tbp]
\begin{center}
\includegraphics[width=\columnwidth]{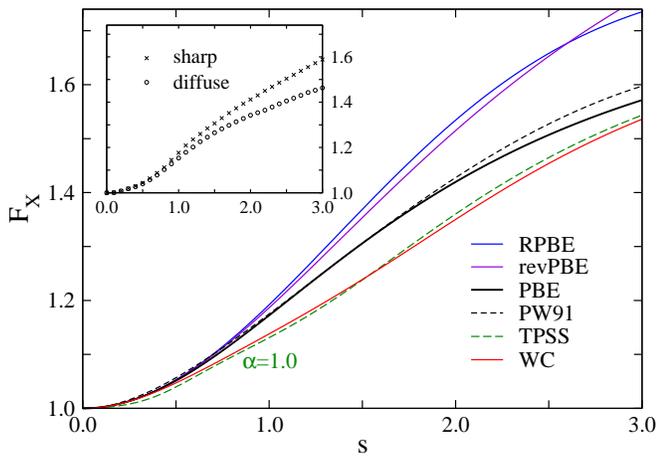}
\end{center}

\caption{\label{fig1} 
(Color online) Exchange enhancement factors $F_{\rm X}$ as functions of 
the reduced gradient $s$. The green curve is $F_{\rm X}$ of TPSS meta-GGA
with $\alpha = 1.0$, which corresponds to slowly varying density limit.
Symbols in the inset are $F_{\rm X}$ determined by the real-space cutoff
procedure. A sharp cutoff generates the crosses, and its parameters are chosen 
to fit $F^{\rm PBE}_{\rm X}$. The circles correspond to a diffuse cutoff
with the same parameters.} 
\end{figure}


Because the magnitude of the exchange energy is much bigger than correlation in most cases, and PBE 
generally describes the correlation energy with enough accuracy, we focus only on the exchange in this 
paper. The dimensionless reduced gradient $s = |\nabla n|/[2(3\pi^{2})^{1/3}n^{4/3}]$. The 
exchange enhancement factor $F_{\rm X}$ is defined as 
\begin{equation}
E^{\rm GGA}_{\rm X} =  
\int n \epsilon_{\rm X}^{\rm unif}(n)F_{\rm X}(s) d^{3}r, 
\label{eq3}
\end{equation}
where the exchange energy density of the uniform electron gas 
$\epsilon_{\rm X}^{\rm unif}(n) = -\frac{3e^2}{4\pi} (3\pi^{2}n)^{1/3}$.
The PBE ensatz of $F_{\rm X}$ has the general form
\begin{equation}
F_{\rm X} = 1 + \kappa - \kappa/(1+x/\kappa),
\label{eq4}
\end{equation}
where $\kappa = 0.804$ to ensure the Lieb-Oxford bound \cite{lieb81}, and $x = \mu s^{2}$ with 
$\mu = 0.21951$ to recover the LSD linear response, i.e., as $s \rightarrow 0$, the
exchange gradient correction cancels that for correlation. In the range of interest 
for real systems $0 \lesssim s \lesssim 3$, the PBE $F_{\rm X}$ is a simple numerical
fit to that of PW91, which is constructed from the gradient expansion of a {\it sharp} real space 
cutoff of the exchange hole \cite{pw91,perdew96}, plus some exact constraints \cite{pbe}. Unlike 
atoms and molecules, solids can have a {\it diffuse} tail around the exchange-correlation hole, and a 
diffuse radial cutoff factor $[1+(u/u_{\rm x})^2]\exp[-(u/u_{\rm x})^2]$,
where $u$ is the distance from the hole center and $u_{\rm x}$ is the fixed radial cutoff, 
leads to a smaller $F_{\rm X}$ \cite{perdew96} for $s \gtrsim 1$ than that 
of the sharp radial cutoff (inset of Figure \ref{fig1}). This explains why the PBE (PW91) functional 
improves total energies of atoms and atomization energies of molecules greatly over LSD, but 
often overcorrects LSD for solids. Two revised versions of PBE, namely revPBE \cite{revpbe} with 
empirical $\kappa = 1.245$ and RPBE \cite{rpbe} with 
$F_{\rm X} = 1 + \kappa - \kappa \exp(-\mu s^{2}/\kappa)$, further exaggerate 
$F_{\rm X}$ (Figure \ref{fig1}), giving better energies for atoms and molecules, 
but worse lattice constants of solids (Table \ref{tab1}).

The real space cutoff procedure can only give qualitative features of $F_{\rm X}$, not the 
exact behavior because it depends on the detailed approximations of the procedure and 
fitting to parameters, so other known constraints must be chosen to determine $F_{\rm X}$. 
Usually valence electron densities of solids vary much more slowly than electron 
densities of atoms and molecules. The choice of $\mu$ in PBE violates the known gradient 
expansion of Svendsen and von Barth \cite{sven} for slowly varying density systems,
\begin{equation}
F_{\rm X} = 1 + \frac{10}{81}p + \frac{146}{2025}q^{2} - \frac{73}{405}qp 
              + Dp^{2} + O(\nabla^{6}),
\label{eq5}
\end{equation}
where $p=s^{2}$, $q=\nabla^{2}n/[4(3\pi^{2})^{2/3}n^{5/3}]$ is the second order reduced 
gradient, and $D=0$ is the best numerical estimate. If $\mu$ is set to $\frac{10}{81}$, 
$F_{\rm X}$ of equation \ref{eq4} will be lowered. 
However the behavior of PBE $F_{\rm X}$ for small $s$ needs
to be retained because (i) it is necessary to retain cancellation of gradient correction of 
exchange and correlation as $s \rightarrow 0$; (ii) $F_{\rm X}$ determined by a diffuse radial 
cutoff is close to that by sharp radial cutoff for small $s$ (inset of Figure \ref{fig1}). 
Thus we propose the following ensatz for $x$ in Equation (\ref{eq4}):
\begin{equation}
x = \frac{10}{81}s^{2} + (\mu - \frac{10}{81})s^{2}\exp(-s^{2}) + \ln (1 + cs^{4}),
\label{eq6}
\end{equation}
where the parameter $c$ is set to recover the fourth order parameters in Equation (\ref{eq5}) for 
small $s$. Because a good approximation of $q$ for slowly varying densities is 
$q \approx \frac{2}{3}p$ \cite{tpss}, 
$c = \frac{146}{2025}(\frac{2}{3})^{2}-\frac{73}{405}\frac{2}{3} + (\mu - \frac{10}{81}) = 0.0079325$.
Our new functional will be referred to as ``WC'', and $F_{\rm X}^{\rm WC}$ 
still satisfies the four conditions (d)-(g) constraining
$F^{\rm PBE}_{\rm X}$ \cite{pbe}. It is nearly identical to $F^{\rm PBE}_{\rm X}$ for 
$s \lesssim 0.5$, and smaller for bigger $s$, as displayed in Figure \ref{fig1}. Amazingly, 
the present simple $F_{\rm X}^{\rm WC}$ matches that of the more byzantine TPSS meta-GGA for 
slowly varying densities very well, which makes use of the kinetic energy density to 
enforce Equation (\ref{eq5}) for small $p$ and $q$. Contrasting with earlier attempts such as 
RevPBE, WC has no adjustable parameters fitted to experimental data, and it is constructed
completely from fundamental physical considerations. 
 
We tested the new functional of Equation (\ref{eq6}) by computing equilibrium crystal structures 
and cohesive energies of solids, jellium surface energies, and exchange energies of atoms. We used 
the planewave pseudopotential method (ABINIT4.4.4 \cite{abinit}) for solids, and an all-electron 
atomic code for atoms. It is straightforward to implement the current GGA from the PBE 
pseudopotential code. We used the same configurations for each atom to generate optimized 
norm-conserving pseudopotentials by the OPIUM code \cite{opium} of LSD, PBE, and WC. 


\begin{table}[tbp]
\caption{\label{tab2}
Errors (mean absolute relative error) of calculated equilibrium lattice constants $a_{0}$, 
bulk moduli $B_{0}$, and cohesive energies $E_{\rm c}$ of 18 tested solids, Li, Na, K, Al, 
C, Si, SiC, Ge, GaAs, NaCl, NaF, LiCl, LiF, MgO, Cu, Rh, Pd, Ag, at 0 K, comparing with 
experiments \cite{star04}. TPSS and PKZB data are from Ref. \cite{star04}. }

\begin{ruledtabular}
\begin{tabular}{lccccc}
      & $a_{0}^{\rm LSD}$  & $a_{0}^{\rm PBE}$ & $a_{0}^{\rm WC}$ 
      & $a_{0}^{\rm TPSS}$ & $a_{0}^{\rm PKZB}$                     \\
Error (\%) &  1.74    & 1.30   &  0.29   & 0.83 &  1.65             \\
\hline
      & $B_{0}^{\rm LSD}$  & $B_{0}^{\rm PBE}$ & $B_{0}^{\rm WC}$ 
      & $B_{0}^{\rm TPSS}$ & $B_{0}^{\rm PKZB}$                     \\
Error (\%) &  12.9    &  9.9   & 3.6     &  7.6  & 8.0              \\
\hline
      & $E_{\rm c}^{\rm LSD}$ & $E_{\rm c}^{\rm PBE}$ & $E_{\rm c}^{\rm WC}$ 
      & &                      \\
Error (\%) &  15.2    &  5.1   & 5.2     &  &               \\

\end{tabular}
\end{ruledtabular}

\end{table}


First we calculated equilibrium lattice constants $a_{0}$ of 18 solids as tested in 
Refs. \cite{tpss,star04}. We found LSD underestimates, while PBE overestimates $a_{0}$, and current 
LSD and PBE errors agree well with previous ones. As seen in Table \ref{tab2}, WC improves $a_{0}$ 
significantly over LSD and PBE, even much better than TPSS. Note that lattice constants should be 
extrapolated to 0 K to compare with DFT results due to thermal expansion. An interesting example 
is $a_{0}$ of cubic PbTiO$_{3}$, which is 3.969 \AA \ at 766 K (ferroelectric phase transition 
temperature). It reduces to 3.93 \AA \ at 0 K by extrapolation \cite{mabud}.
$a^{\rm PBE}_{0} = 3.971$ \AA \ is 1\% larger, whereas $a^{\rm WC}_{0} = 3.933$ \AA, in 
excellent agreement with the extrapolated data. Also note that the zero-point quantum 
fluctuations are not included in DFT calculations, which would expand $a_{0}$ about 0.2\%. WC also 
predicts more accurate bulk moduli for these materials than LSD and PBE (Table \ref{tab2}). For
cohesive energies, WC is nearly as accurate as PBE, and much better than LSD (Table \ref{tab2}).  
These results prove that our simple model of GGA is very suitable for solids.


\begin{table}[tbp]
\caption{\label{tab3} 
Equilibrium structural parameters for two ferroelectrics: tetragonal
$P4mm$ PbTiO$_{3}$ (PT) and rhombohedral $R3m$ BaTiO$_{3}$ (BT). The atom 
positions $u_{z}$ are given in terms of the lattice constants.}

\begin{ruledtabular}
\begin{tabular}{llcccc}
            &                     & LSD    & PBE    & WC     & Expt.  \\
\hline
PT          & $V_{0}$ (\AA$^{3}$) & 60.37  & 70.54  & 63.47  & 63.09\footnotemark[1]  \\
            & $c/a$               & 1.046  & 1.239  & 1.078  & 1.071\footnotemark[1]  \\
            & $u_{z}$(Pb)         & 0.0000 & 0.0000 & 0.0000 & 0.000\footnotemark[2]  \\  
            & $u_{z}$(Ti)         & 0.5235 & 0.5532 & 0.5324 & 0.538\footnotemark[2]  \\  
            & $u_{z}$(O$_1$O$_1$) & 0.5886 & 0.6615 & 0.6106 & 0.612\footnotemark[2]  \\  
            & $u_{z}$(O$_3$)      & 0.0823 & 0.1884 & 0.1083 & 0.112\footnotemark[2]  \\  
\hline
BT          & $V_{0}$ (\AA$^{3}$) & 61.59  & 67.47  & 64.04  & 64.04\footnotemark[3]  \\
            & $\alpha$            & 89.91$^{\circ}$ & 89.65$^{\circ}$\footnotemark[3]  
                                  & 89.86$^{\circ}$ & 89.87$^{\circ}$\footnotemark[3] \\
            & $u_{z}$(Ba)         & 0.0000 & 0.0000 & 0.0000 & 0.000\footnotemark[3]  \\  
            & $u_{z}$(Ti)         & 0.4901 & 0.4845 & 0.4883 & 0.487\footnotemark[3]  \\  
            & $u_{z}$(O$_1$O$_1$) & 0.5092 & 0.5172 & 0.5116 & 0.511\footnotemark[3]  \\  
            & $u_{z}$(O$_3$)      & 0.0150 & 0.0295 & 0.0184 & 0.018\footnotemark[3]  \\  

\end{tabular}
\end{ruledtabular}
\footnotetext[1]{Low temperature data, Ref. \cite{mabud}.}
\footnotetext[2]{Room temperature data, Ref. \cite{pt}.}
\footnotetext[3]{Low temperature data, Ref. \cite{bt}.}

\end{table}


For the ground-state structures of polarized ferroelectrics, the lattice strain must be optimized
together with atomic positions. Unlike cubic systems, a large volume for a polarized ferroelectric 
material favors large strain and atomic displacements, and large strain and atomic displacements 
lead to even larger volumes. This causes PBE to overestimate the volume of tetragonal PbTiO$_{3}$ 
by more than 10\%, whereas the error is only 3\% for the cubic structure. Table \ref{tab3} 
summarizes the LSD, PBE, and WC results of fully relaxed tetragonal PbTiO$_{3}$ and rhombohedral 
BaTiO$_{3}$. It shows that WC predicts highly accurate volumes, strains, and atomic 
displacements, whereas LSD and PBE underestimate and overestimate these values, respectively. If 
their model parameters are fitted to first-principles results using WC, MD or MC simulations are 
expected to determine ferroelectric phase transition temperatures and other properties more accurately. 
 
It is well known that LSD fails to predict the correct ground states for certain materials,
e.g., magnetic bcc iron \cite{iron} and $\alpha$-quartz \cite{quartz}, and PBE can 
eliminate this error. WC predicts correct ground states for both iron and quartz with
smaller energy differences than PBE, resulting in lower transition pressures.
For iron, the transition (bcc to hcp) pressure of 10 GPa is rather close to experiment, but for 
quartz ($\alpha$ to stishovite), the WC result of 2.6 GPa is not sufficient to 
correct LSD. Since the stishovite phase is much more compact and stiffer than the $\alpha$ phase,
the phonon contributions to energy could increase the energy difference greatly. 
However by performing first-principles linear response lattice dynamics calculations we find that 
the vibration zero point energy difference is only 0.015 eV/SiO$_2$,
because the $\alpha$ phase has high frequency modes the stishovite phase lacks,
in addition to low frequency modes. These results indicate that a better correlation 
functional for WC is also needed.


\begin{figure}[tbp]
\begin{center}
\includegraphics[width=\columnwidth]{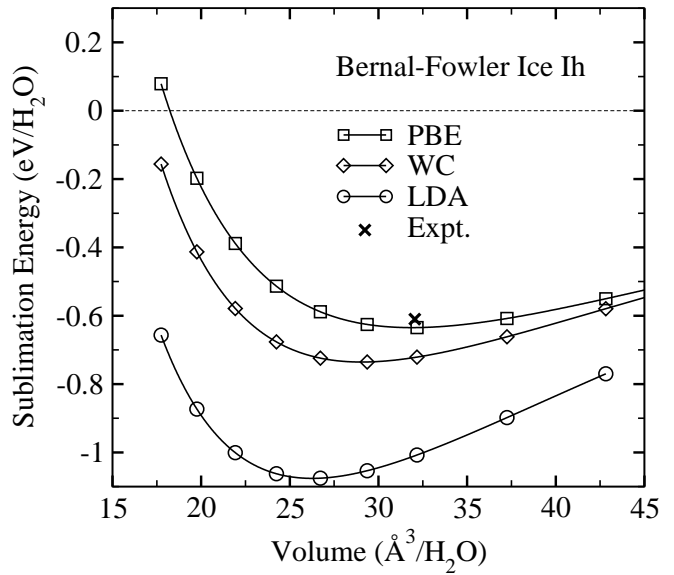}
\end{center}

\caption{\label{fig2} 
Sublimation energies of Bernal-Fowler periodic model of ice Ih.}
\end{figure}


Another interesting case is the weakly interacting molecular bonding systems. We calculated the 
hydrogen bond strength in ice for a periodic Bernal-Fowler Ih model \cite{ice1,ice2}. 
The WC sublimation energy of 0.73 eV per H$_2$O (the measured data is 0.61 eV \cite{ice3}, excluding 
the zero-point vibration) is much better than LSD of 1.07 eV, but not as good as PBE of 0.63 eV, 
as shown in Figure \ref{fig2}. 


\begin{table}
\caption{\label{tab4}
Jellium surface exchange energies ($\sigma_{\rm X}$, in erg/cm$^2$) computed using the 
LSD orbitals and densities. LSD and PBE values are from Ref. \cite{pkzb}, and
TPSS values are from Ref. \cite{star04}. The last row is the mean absolute relative errors 
against the exact value \cite{pitarke}. }

\begin{ruledtabular}
\begin{tabular}{lccccc}
$r_s$ (bohr) &    LSD   &    PBE   &    TPSS  &    WC   & exact       \\
\hline
2.00         &   3037   &   2438   &    2553  &  2519   &  2624       \\
2.30         &	 1809   &   1395   &    1469  &  1452	&  1521       \\
2.66         &   1051	&    770   &     817  &   809   &   854       \\
3.00         &    669   &    468   &     497  &   497   &   526       \\
3.28         &	  477   &    318   &     341  &   341   &   364       \\
4.00         &    222   &    128   &     141  &   141   &   157       \\
5.00         &	   92	&     40   &      47  &    47	&    57       \\
6.00         &	   43	&     12   &      15  &    15	&    22       \\ 

Error (\%)   &   36.7   &   16.7   &     9.4  &   9.8   &             \\
\end{tabular}
\end{ruledtabular}

\end{table}


As argued by Perdew \cite{dft2}, in an extended system such as metal surface the exact hole may
display a diffuse long-tail behavior: an emitted electron's exchange-correlation hole can extend 
significantly back into the interior of the metal. As summarized in Table \ref{tab4}, jellium surface
exchange energy $\sigma_{\rm X}$ is severely overestimated by LSD and underestimated by PBE, and
$\sigma_{\rm X}^{\rm WC}$ are better than both $\sigma_{\rm X}^{\rm LSD}$ and 
$\sigma_{\rm X}^{\rm PBE}$, identical to $\sigma_{\rm X}^{\rm TPSS}$ for 
$r_{\rm s} \geq 3.0$ bohr, where $r_{\rm s} = (\frac{3}{4\pi n})^{1/3}$. 
Because the correlation surface energy $\sigma_{\rm C}^{\rm TPSS}$ is 
very close to $\sigma_{\rm C}^{\rm PBE}$, the mean error of $\sigma^{\rm WC}_{\rm XC}$
against the available most accurate $\sigma_{\rm XC}^{\rm RPA+}$ \cite{yan00}, which is 1.5\%, is 
comparable to 1.1\% of TPSS, better than the LSD and PBE errors of $\sigma_{\rm XC}$ of 2.1\% 
and 4.9\%, respectively. The significant improvement of WC upon PBE for the jellium 
surface energy is because the diffuse radial cutoff model better describes metal surfaces than 
the sharp cutoff. The good performance of WC on jellium surface energies suggests 
it should also perform well on metal vacancies \cite{anne}.
  
Finally we compared calculated exchange energies of 5 noble-gas atoms with the
Hartree-Fock (HF) results \cite{perdew92}. The mean errors for LSD,
PBE, RPBE, and WC are 8.77\%, 0.89\%, 0.18\%, and 2.01\%, respectively.
Comparing the magnitude of $F_{\rm X}$ of these GGAs as illustrated in Figure \ref{fig1},
one can conclude that among these choices a GGA with bigger $F_{\rm X}$ predicts better
$E_{\rm X}$ of atoms. Although WC is constructed for slowly varying densities, it improves 
exchange energies of atoms over LSD significantly. Furthermore, a proper treatment of correlation
functional would make WC XC energies of atoms as accurate as PBE. 

We have shown that exchange enhancement factors $F_{\rm X}$ constructed from different situations
perform with different accuracy for the same system. Using PBE correlation functional,
WC performs excellently for solids, but it is less accurate for atoms than PBE; 
on the other hand, RPBE is excellent for atoms, but poorer for solids than PBE. 
It seems impossible to make a GGA which is more accurate than PBE for both solids 
and atoms simultaneously because $F_{\rm X}$ is a function {\it only} of the reduced gradient 
$s$. Our calculations show that $F_{\rm X}$ also depends on the variation of $|\nabla n|$. Since 
high density systems often have large variations and low density systems often vary slowly, we 
propose that a GGA having $F_{\rm X} = F_{\rm X}(s, r_{\rm s})$, just like the gradient correlation 
correction, could be universally more accurate for atoms, molecules, and solids than PBE. 
The additional parameters of $F_{\rm X}(s, r_{\rm s})$ can be fitted to quantum Monte Carlo 
simulations for specific materials, or determined by constraints at 
$r_{\rm s} \rightarrow 0$ and $r_{\rm s} \rightarrow \infty$ from other theoretical considerations. 
In addition, a better correlation functional will improve the XC energy and potential also, and 
one can construct a correlation more compatible with our functional of exchange than PBE.
In this way, the accuracy of a simple second rung of the ladder of XC approximations, GGA, could 
approach that of the more complicated third rung approximation, meta-GGA.

We have constructed a new GGA which is more accurate for solids than any existing GGA
and meta-GGA. It has a very simple form without any empirical parameters, and it is ideal 
for {\it ab inito} calculations of certain materials, e.g., ferroelectrics, for which 
exceptionally high accuracy is needed. It can be generalized to make a GGA more accurate for 
atoms, molecules, and solids than PBE.

We are indebted to L. Almeida and J. P. Perdew for sending us the jellium code. We thank 
E. J. Walter, H. Krakauer, P. Schultz, and A. E. Mattsson for helpful discussions. 
This work was supported by the Center for Piezoelectrics by Design (CPD) and the Office 
of Naval Research (ONR) under ONR Grants No. N00014-02-1-0506.



\begin{references}

\bibitem{dft1}
W. Kohn and L. J. Sham, Phys. Rev. {\bf 140}, A1133 (1965).

\bibitem{dft2}
{\it A Primer in Density Functional Theory}, edited by C. Fiolhais, F. Nogueira, and M. Marques 
(Springer, Berlin, 2003).

\bibitem{pw86}
J. P. Perdew and Y. Wang, \prb {\bf 33}, R8800 (1986).

\bibitem{pw91}
J. P. Perdew, in {\it Electronic Structure of Solids '91}, 
edited by P. Ziesche and H. Eschrig (Akademie Verlag, Berlin, 1991).

\bibitem{pbe}
J. P. Perdew, K. Burke, and M. Ernzerhof, Phys. Rev. Lett. {\bf 77}, 3865 (1996).

\bibitem{perdew92}
J. P. Perdew, {\it et al.}, \prb {\bf 46}, 6671 (1992).

\bibitem{cohen90}
R. E. Cohen and H. Krakauer, Phys. Rev. B {\bf 42}, 6416 (1990).

\bibitem{cohen92}
R. E. Cohen, Nature (London) {\bf 358}, 136 (1992).

\bibitem{singh92}
D. J. Singh and L. L. Boyer, Ferroelectrics {\bf 136}, 95 (1992).

\bibitem{star04}
V. N. Staroverov, {\it et al.}, \prb {\bf 69}, 075102 (2004).

\bibitem{tpss}
J. Tao, {\it et al.}, \prl {\bf 91}, 146401 (2003).

\bibitem{pkzb}
J. P. Perdew, {\it et al.}, \prl {\bf 82}, 2544 (1999).

\bibitem{wu04}
Z. Wu, R. E. Cohen, and D. J. Singh, \prb {\bf 70}, 104112 (2004).

\bibitem{revpbe}
Y. Zhang and W. Yang, \prl {\bf 80}, 890 (1998).

\bibitem{rpbe}
B. Hammer, L. B. Hansen, and J. K. N{\o}rskov, \prb {\bf 59}, 7413 (1999).

\bibitem{zhong94}
W. Zhong, D. Vanderbilt, and K. M. Rabe, \prl {\bf 73}, 1861 (1994);
\prb {\bf 52}, 6301 (1995).

\bibitem{marcelo04}
M. Sepliarsky, Z. Wu, and R. E. Cohen, unpublished;
M. Sepliarsky, Z. Wu, A. Asthagiri, and R. E. Cohen, Ferroelectrics {\bf 301}, 55 (2004).

\bibitem{krakauer99}
H. Krakauer, {\it et al.}, J. Phys.: Condens. Matter {\bf 11}, 3779 (1999).

\bibitem{lieb81}
E. H. Lieb and S. Oxford, Int. J. Quantum Chem. {\bf 19}, 427 (1981).

\bibitem{perdew96}
J. P. Perdew, K. Burke, and Yue Wang, Phys. Rev. B {\bf 54}, 16533 (1996).

\bibitem{sven}
P. S. Svendsen and U. von Barth, \prb {\bf 54}, 17402 (1996).

\bibitem{abinit}
X. Gonze, {\it et al.}, Comput. Mater. Sci. {\bf 25}, 478 (2002);
http://www.abinit.org. 

\bibitem{opium} 
A. M. Rappe, {\it et al.}, Phys. Rev. B {\bf 41}, R1227 (1990);
http://opium.sourceforge.net.

\bibitem{mabud}
S. A. Mabud and A. M. Glazer, J. Appl. Cryst. {\bf 12}, 49 (1979).

\bibitem{exx}
M. St\"{a}dele, {\it et al.}, \prb {\bf 59}, 10031 (1999). 

\bibitem{pt}
G. Shirane, {\it et al.}, Acta Cryst, {\bf 9} 131 (1956).

\bibitem{bt}
A. H. Hewat, Ferroelectrics, {\bf 6}, 215 (1074).

\bibitem{iron}
L. Stixrude, R. E. Cohen, and D. J. Singh, \prb {\bf 50}, 6442 (1994).

\bibitem{quartz}
D. R. Hamann, \prl {\bf 76}, 660 (1996).

\bibitem{ice1}
D. R. Hamann, \prb {\bf 55}, R10157 (1997).

\bibitem{ice2}
P. J. Feibelman, Science, {\bf 295}, 99 (2002)

\bibitem{ice3}
E. Whalley, in {\it The Hydrogen Bond}, P. Schuster, G. Zundel, C. Sandorfy, Eds. 
(North-Holland, Amsterdam, 1976), vol. 3, pp.1425-1470. According to his analysis, zero-point 
vibration reduces the 0 K sublimation energy of H$_{2}$O by 120 meV and of D$_{2}$O ice by 98 meV.

\bibitem{yan00}
Z. Yan, J. P. Perdew, and S. Kurth, \prb {\bf 61}, 16430 (2000).

\bibitem{pitarke}
J. M. Pitarke and A. G. Eguiluz, \prb {\bf 63}, 045116 (2001).

\bibitem{anne}
A. E. Mattsson, private communication.

\end{references}
\end{document}